\documentclass[twocolumn,showpacs,aps,prb]{revtex4}

\usepackage{dcolumn}
\usepackage{epsfig}
\def\gsim{\; $\raise0.3ex\hbox{$>$}\llap{\lower0.8ex\hbox{$\sim$}}$\;}
\def\lsim{\; $\raise0.3ex\hbox{$<$}\llap{\lower0.8ex\hbox{$\sim$}}$\;}

\begin{document}
\title{Filling-dependence of the zigzag Hubbard ladder for a quasi-one-dimensional superconductor Pr$_2$Ba$_4$Cu$_7$O$_{15-\delta} $ }
\author{Kouichi Okunishi}
\affiliation{
Department of Physics, Faculty of Science, Niigata University, Igarashi 2, 950-2181, Japan
}
\date{\today}

\begin{abstract}
We investigate  filling dependence of the zigzag Hubbard ladder, using density matrix renormalization group method. 
We illustrate the chemical-potential vs. electron-density and spin gap vs. electron density curves, which reflect characteristic properties of the electron state.
On the basis of the obtained phase diagram, we discuss the connection to a novel quasi-one-dimensional superconductor Pr$_2$Ba$_4$Cu$_7$O$_{15-\delta}$.
\end{abstract}


\maketitle

\section{introduction}

Low-dimensional strongly-correlated-electron systems have been one of the most active research fields in the condensed matter physics, since the discovery of high-$T_c$ superconductivity.\cite{hightc}
One of the current topics in the field is the frustration effect, which often causes a variety of interesting behavior such as exotic superconductivity.
Recently, a novel quasi-one-dimensional(1D) compound Pr$_2$Ba$_4$Cu$_7$O$_{15-\delta}$(Pr247) was found to exhibit the superconductivity below $T_c\simeq 18$K. This compound has the layered structure of single chains, frustrated zigzag ladders, and CuO$_2$ planes.\cite{matsukawa,yamada}
An important  point on Pr247 is that the metallic conductivity originates from the zigzag ladder part, while the CuO$_2$ plane is insulating with the antiferromagnetic order below 285K.\cite{rice,shinji2, shinji}
Moreover the oxygen atoms at the single chains, which are also insulating,  are easily defected with the deoxidization and then the electrons are effectively doped into the zigzag ladders from the single chain sites.
This suggests that  Pr247 can be regarded as a natural ``filling controlled system'', which hopefully provides an interesting physics cooperatively induced by the filling dependence and the frustration effect.
In fact, the transition temperature of Pr247 exhibits the  systematical relation  to the oxygen deficiency $\delta$, implying that the electron-filling plays a crucial role for appearance of the superconductivity.

A relevant model describing Pr247 is the zigzag Hubbard ladder, whose Hamiltonian is given by
\begin{eqnarray}
{\cal H}= \sum_{i\sigma}[t_1 c^\dagger_{i\sigma} c_{i+1\sigma}+ t_2  c^\dagger_{i\sigma} c_{i+2\sigma}+ h.c. ] + U \sum_i n_{i\uparrow}n_{i\downarrow},
\label{zigzag}
\end{eqnarray}
where $c_{i\sigma}$ is the electron annihilation operator at $i$ site with spin $\sigma$ and  $n_{i\sigma}\equiv c^\dagger_{i\sigma}c_{i\sigma}$.\cite{sano}
$t_1$($t_2$) is the nearest(next-nearest) hopping term of the electron and $U$ is the on site coulomb energy.
We also introduce $\alpha=|t_1/t_2|$ for later convenience.
According to Ref.\cite{sano}, the  overlap of Cu-$3d$ orbit and O-$2p$ orbit in the zigzag array of Cu-O double chain structure suggests that the parameters corresponding to Pr247 are in  $t_2<0$ and $ \alpha <1 $.
We also assume  $U= 8|t_2|$ or $8|t_1|$, which may be of similar order to the usual cuprates.
The filling of the zigzag ladder part in Pr247 of $\delta=0$ corresponds to the nearly quarter filling.\cite{shinji2}
As the oxygen deficiency $\delta$ increases, it  rises continuously toward half-filled side and the  superconductivity appears for  $\delta > 0.3 $.
This indicates that it is primarily important to investigate  the frustration and  filling dependences of the electron state  systematically,  for through understanding of the Pr247 superconductivity.

The Hamiltonian (\ref{zigzag}) is a typical example of the 1D correlated electron system capturing various frustration effects.
Indeed, interesting properties of low energy excitations are revealed at the half-filling or quarter filling  by extensive studies with bosonization and various numerical methods.\cite{fabrizio,kuroki,arita,daul,daul2,torio,japar}
In particular, Fabrizio sketched the qualitative phase diagram\cite{fabrizio} by invoking the weak coupling theory for the non-frustrating ladder system\cite{balentz},  and Daul etal presented an approximate phase diagram  in the context of the ferromagnetism\cite{daul}, which also  reveal that the zigzag Hubbard model contains  quite rich physics.
However, both the frustration and the incommensurate Fermi wave number make a  quantitative analysis of  the low-energy excitations subtle;
the precise filling-dependence of the spin gap is still unclear in the relevant parameter region to Pr247.

In this paper, we precisely investigate the filling dependence of the zigzag Hubbard ladder with intensive calculations of density matrix renormalization group(DMRG)\cite{white}.
In the next section we particulary illustrate the chemical potential-electron density($\mu$-$\rho$) curve.
In section III, we investigate the filling dependence of the spin gap in the region $t_2<0$, from which we  read characteristic properties of the electron state .
In sections IV and V, we summarize the  DMRG results as a phase diagram and then discuss the relevance to the superconductivity in Pr247 respectively.
Here, we note that the convergence of DMRG computation in the single band region is good with a relatively small number of retained bases $m\sim 200$, while in the two band region, we can obtain the reliable spin gap with up to $m=1000$.

\section{$\mu$-$\rho$ curves}

Let us write the gourd state energy  of $L$ sites as $E_L(N,S^z)$, where  $N$ denotes the number of electrons and $S^z$ indicates the total-$S^z$ of the system.
The electron density is written as $\rho\equiv N/L$ and then the chemical potential is defined as  $\mu= -[E_L(N+ 1,1/2)-E_L(N,0)]$ for $N$=even or  $\mu= -[E_L(N+1,0)-E_L(N,1/2)]$ for $N$=odd.
This definition of $\mu$ should be contrasted to the conventional notation of even number of electrons.
This is because the $\mu$-$\rho$ curve of the present definition of $\mu$ can visualize some important natures of the low-energy excitation;
If the charge excitation is gapless, the $\mu$-$\rho$ curve acquires a smooth stairway-like curve. 
If two electrons conform a bound state, the  $N=$even case is slightly stable than the $N=$odd case, due to the binding energy of the electrons.
This suggests that if the system has the bound state, the $\mu$-$\rho$ curve exhibits overhung behavior. 
We should  note that the similar behavior can be actually seen in the magnetization curve of the zigzag spin system.\cite{zigzagspin}

As can be expected, the basic property of the $\mu$-$\rho$ curve is attributed to the one-particle dispersion curve that is easily obtained as
\begin{equation}
\varepsilon(k)=2t_1 \cos(k)+2t_2 \cos(2k) .\label{disp1}
\end{equation}
The shape of this dispersion curve has double-well form for $\alpha < 4$(recall the sign of $t_2$ is negative), where it has the van Hove singularity.
In the following, the corresponding electron density is denoted  as $\rho_c$.
As $\alpha$ increases, $\rho_c$ comes down to the lower-electron density region.
For $ \alpha < 2$, $\rho_c$ reaches below the half filling($\rho<1$), and, at $\alpha= 1/2$, it is located near the quarter filling.
As $\alpha\to 0$, the system finally becomes the  two decoupled Hubbard chains. 
Thus the system in the non-interacting limit($U=0$) is essentially a single band system for $\rho < \rho_{c}$, while it is a two band system having four Fermi points for $ \rho > \rho_{c}$.
We note that a finite $U$ may induce the spin gap in the two band region, according to the weak coupling theory.

\begin{figure}[tb]
\epsfig{file=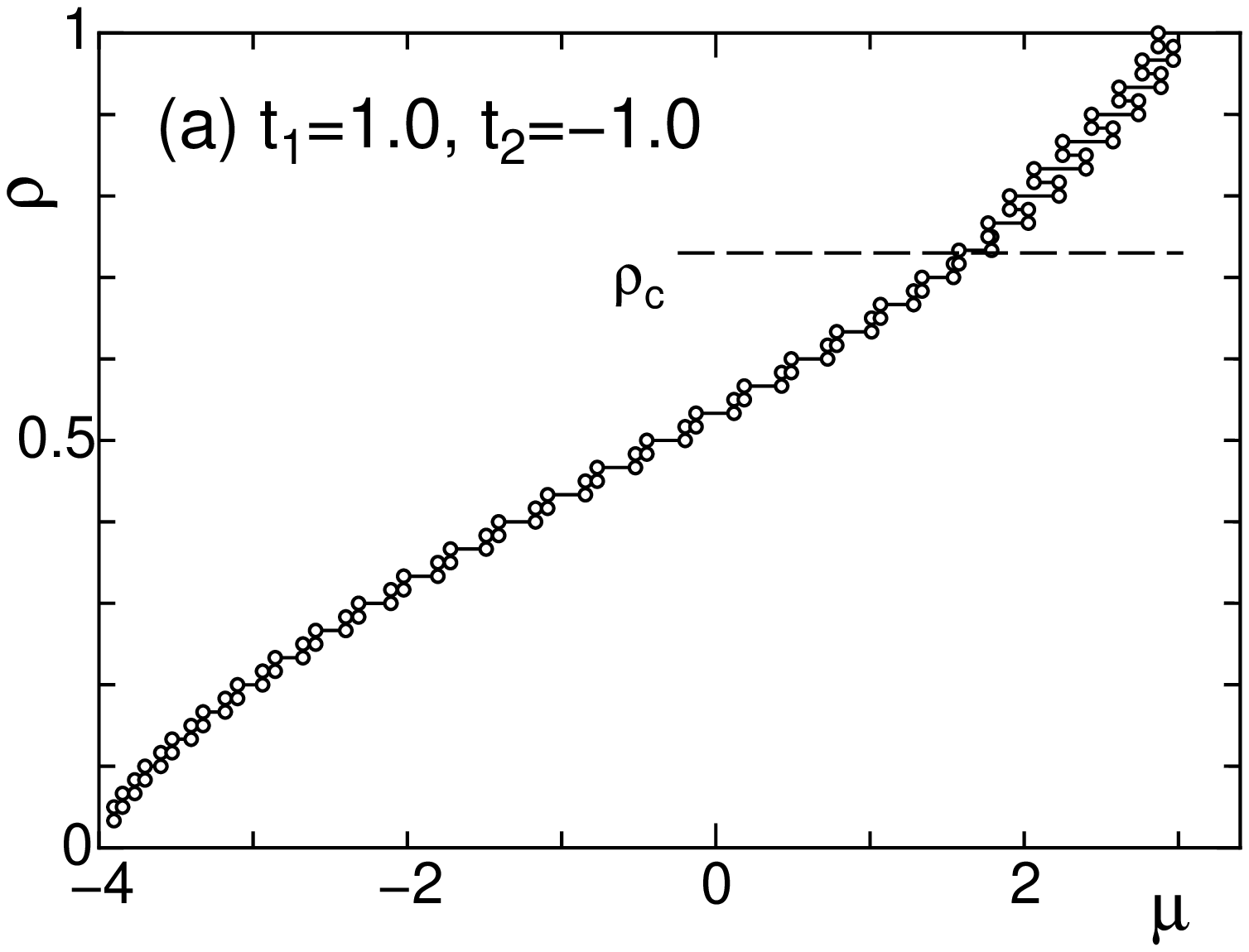,width=7cm}
\epsfig{file=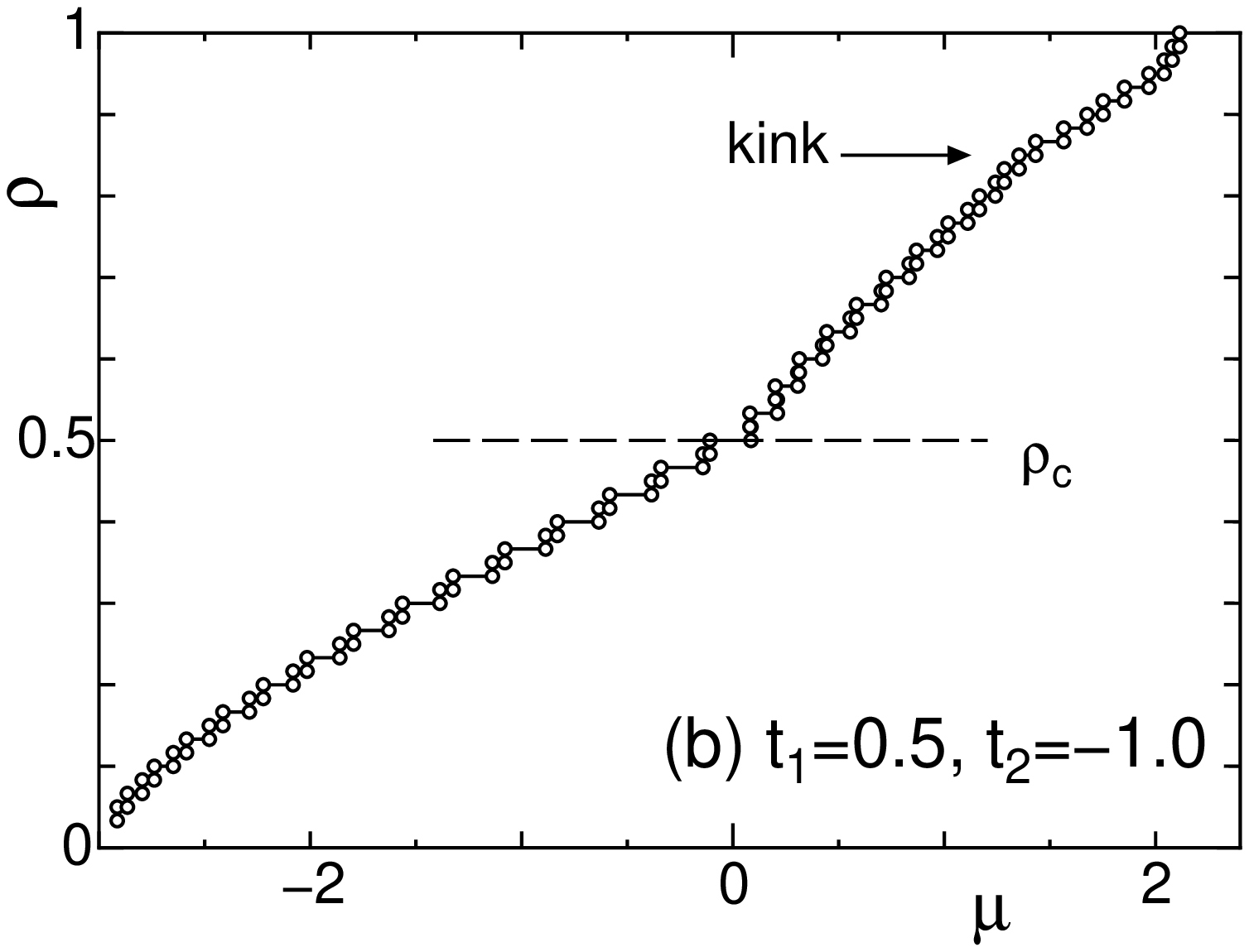,width=7cm}
\caption{$\mu$-$\rho$ curves for $L=60$ systems of $U=8$:
(a) $t_1=1$ and $t_2=-1$($\alpha=1$) and (b) $t_1=0.5$ and $t_2=-1$($\alpha=0.5$). 
The broken lines are guides for $\rho_c$.}
\label{mr}
\end{figure}

In Fig. \ref{mr}, we show the  $\mu$-$\rho$ curves of $U=8$ for $\alpha=1$ and $\alpha=0.5$ with $t_2=-1$.
In the figure, we can find a characteristic feature:
The curvature of the  $\mu$-$\rho$ curve changes at $\rho_c\simeq 0.73$ for $\alpha=1$ and $\rho_c\simeq 0.5$ for $\alpha=0.5$, which is basically reflecting the shape of the dispersion curve (\ref{disp1}).
We however note that these values of $\rho_c$ are slightly lifted from the non-interacting case due to the effect of $U$.
A more interesting point is that  the  $\mu$-$\rho$ curves actually exhibits the oscillating behavior above $\rho_c$.
For $\alpha=1$, we can see the overhung behavior in $\rho_c <\rho < 1$, implying that  the spin gap may exist between $\rho_c$ and the half filling.
On the other hand, for $\alpha =0.5$, the oscillation is enhanced  above $\rho_c$, but the $\mu$-$\rho$ curve becomes smooth again above $\rho\simeq 0.83$ as is indicated by an arrow in Fig.\ref{mr} (b).
This suggests that the spin gap for $\alpha=0.5$ may appear only near $\rho_c$.
Here, we should note that the even-odd effect below $\rho_c$ is properly removed by the size extrapolation.

\section{spin gap}

Since we have seen the outline of the filling dependence of the system in the $\mu$-$\rho$ curve, we further  analyze the spin gap for the precise characterization of the low-energy excitation. 
We define the spin gap $\Delta_s$ as
\begin{eqnarray}
\Delta_s(N;L) &=& E_L(N,1)-E_L(N,0),
\end{eqnarray}
for $N=$even.
In order to extract the bulk behavior of the spin gap, the finite size extrapolation is usually required.
We should however recall that $\rho$ takes only some fractional values restricted by combination of  $L$ and $N$, as far as we treat a finite size system.
In order to avoid this mismatching problem of the electron filling, we calculate $\Delta_s$ of all the electron numbers for various $L$, with which we interpolate the spin gap at an irrational value of $\rho$.
We think that the bulk spin gap can be properly analyzed except for the vicinities of some singular points.

\begin{figure}[tb]
\epsfig{file=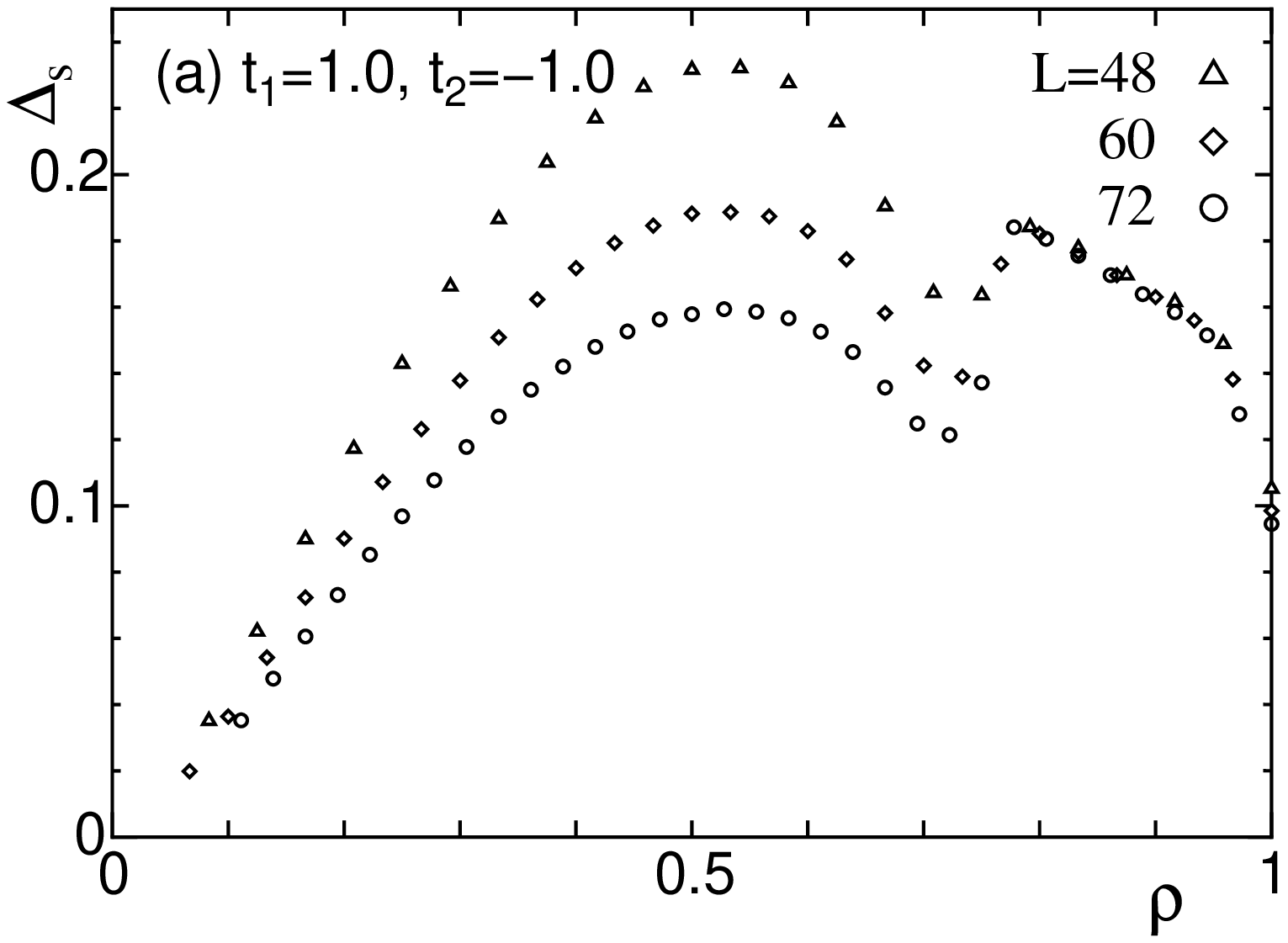,width=7cm}
\epsfig{file=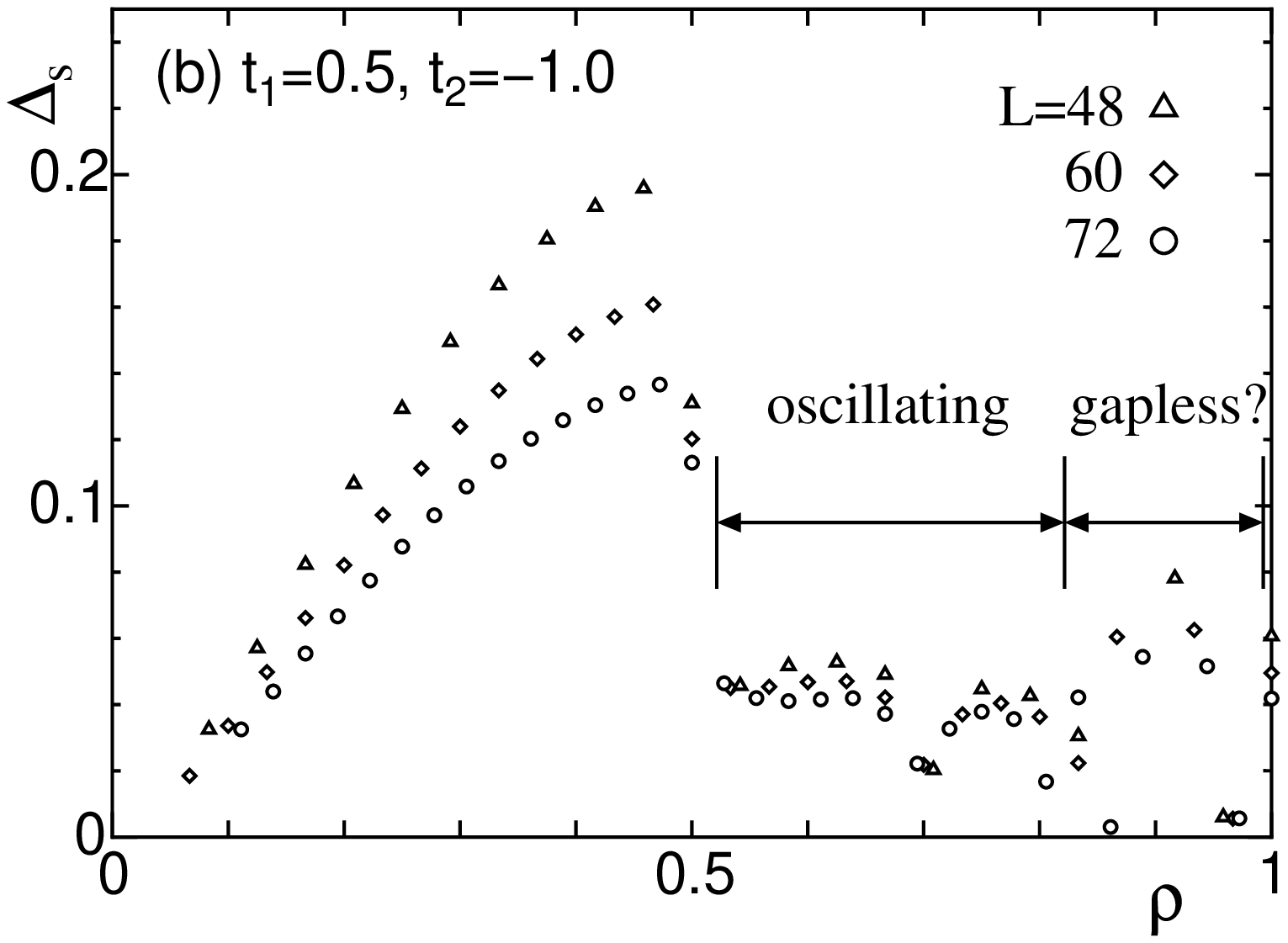,width=7cm}
\caption{Size dependence of spin gaps for $U=8$: (a) $t_1=1$ and $t_2=-1$ and (b) $t_1=0.5$ and $t_2=-1$.}
\label{sgap}
\end{figure}

In Fig. \ref{sgap} (a), we show the DMRG calculated spin gap for $L=48, 60, 72$ with  $t_1=-t_2=1$($\alpha=1$).
We can see that the property of the spin gap clearly changes at $\rho_c\simeq 0.7$, which is consistent with the $\mu$-$\rho$ curve.
Below $\rho_c$, the system is essentially described by the single band and $\Delta_s$ shows large size dependence, suggesting that the spin excitation is gapless.
We thus examine the $1/L$-size dependence of the interpolated spin gap: $\Delta_s(\rho;L)= \Delta_s(\rho) + {\rm const}/L$,  and then verify that $\Delta_s$ below $\rho_c$ becomes zero in $L\to \infty$.
On the other hand, we can expect the spin gap in $\rho_c < \rho <1$, where the overhung behavior of the $\mu$-$\rho$ curves is observed.
In fact we can clearly see that  $\Delta_s$ is almost independent of the system size,  implying that the spin gap remains a finite value in the bulk limit.
Moreover, we can find that the spin gap is enhanced in $\rho< 1$ rather than the half-filling, which may be a peculiar behavior in the frustrating system in contrast to the non-frustrating Hubbard ladder.
Here, we note that this spin gap phase is adiabatically connected to the dimer spin gap of the corresponding zigzag spin system at the half filling.
The amplitude of the spin gap at the half filling is consistent with the zigzag spin chain\cite{whiteaffleck}.

Fig. \ref{sgap} (b) shows $\Delta_s$ for $L=$ 48, 60, 72 with  $t_1=0.5$ and  $t_2=-1$($\alpha=0.5$).
For $\rho<\rho_c(\simeq 0.5)$, the system is essentially a  single band model and thus the size dependence of $\Delta_s$ is basically the same as that for $\alpha=1$;
the $1/L$ extrapolation of $\Delta_s$ leads that the spin excitation is gapless in the thermodynamic limit.
On the other hand, $\Delta_s$ above $\rho_c$ shows subtle behaviors.
Here, we should recall that the $\mu$-$\rho$ curve for $\alpha=0.5$ has a weak anomaly at $\rho\simeq 0.83$, above which  $\mu$-$\rho$ becomes smooth. 
We first analyze the region of $\rho_c< \rho < 0.83 $, for which the even-odd oscillation appears in the $\mu$-$\rho$ curve.
In this region, the spin gap shows oscillating behavior with respect to $\rho$ and $\alpha$.
The amplitude of the oscillation increases, as $\rho$ increases from $\rho_c$ to 0.83.
Thus the precise extrapolation of $\Delta_s$ is still difficult in this region.
However, we note that the size dependence of $\Delta_s$ in the vicinity of $\rho_c$($\rho_c < \rho <0.7$) is rather weak and  a small spin gap may survive in the bulk limit.

Next let us turn to $0.83 < \rho <1$, where we can see the similar oscillating behavior of $\Delta_s$.
However, this oscillation  shows rather systematical behavior depending  on $N\; {\rm mod}\; 4 = 0$ or $\ne 0$. 
In this region,  the system is basically described by the two chains, where the electron filling is sufficiently away from $\rho_c$.
Since the system has four Fermi points,  the $N\; {\rm mod}\; 4 = 0$ state  is stabilized by conforming ``closed shell'' with respect to the  four Fermi points, while the $N\; {\rm mod}\; 4 \ne 0$ state has  a ``unoccupied orbits'', which may generate an anomalous spin excitation. 
We thus read the bulk behavior of the spin gap from  $N\ {\rm mod}\; 4 = 0$ sectors, which shows large size dependence suggesting that the spin excitation seems to become gapless or very small. 
However, the precise estimation of the spin gap value is also difficult within  $L=72$.

Here, we make a comment on the boundary effect, since a boundary excitation may appear in the gapped frustrating system with the open boundary.
In order to check it, we have also calculated magnetization curve at some typical fillings.
Then we can verify that the $S^z=1$ excitation smoothly  connects to the bulk part of the magnetization curve in the spin gap phase of $\alpha=1$.
On the other hand, the magnetization curve of  $\alpha=0.5$ suggests that the $\Delta_s$ sometimes capture the boundary effect in the oscillating spin gap region, where the $S^z=1$ excitation sometimes generates an anomalous step away from the bulk part of the magnetization curve.

\section{phase diagram}

\begin{figure}[bt]
\epsfig{file=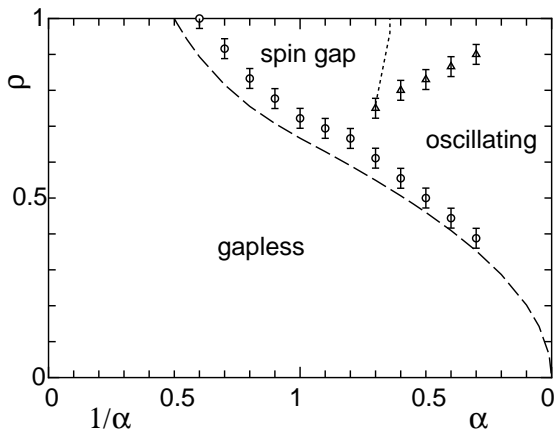,width=7.5cm}
\caption{Spin-excitation phase diagram of the zigzag Hubbard ladder of $U=8$.
In the left-half,  $0< -t_2 <1$ with $t_1=1$, and in the right half, $ 0< t_1<1 $ with $t_2=-1$.
Open circles indicate the boundary between the single-band gapless phase and the spin gap phase. The dotted line is a guide for the vanishing line of the  spin gap.
Triangles mean the boundary between the (almost)gapless region and the oscillating spin gap region. 
The broken line means $\rho_c$ of the free electron case.
The error bar corresponds to the interval of descritized  density of electron for a $L=72$ system.
}
\label{phase}
\end{figure}

In Fig. \ref{phase}, we summarize the phase diagram of the zigzag Hubbard model for $U=8$, which is determined by  the $\mu$-$\rho$ curves and the spin gap for $L=72$.
The open circles indicate the boundary between the single-band gapless phase and spin gap/oscillating region, which is slightly lifted from $\rho_c$ for the free electron case due to the electron correlation effect.
The `` spin gap'' indicates the spin gap phase, which is  connected to the dimer spin gap phase at the half filling\cite{kuroki,torio}.
As $\alpha$ decreases from the dimer spin gap phase, the spin gap rapidly decreases  and  almost vanishes across the dotted line.
We  also note that spin gap at the half-filling  for  $\alpha <0.7$ seems to be almost gapless.
However, it is still difficult to  distinguish whether the spin excitation in this regime is the truly gapless phase or very small spin gap phase within the present accuracy.

\begin{figure}[bt]
\epsfig{file=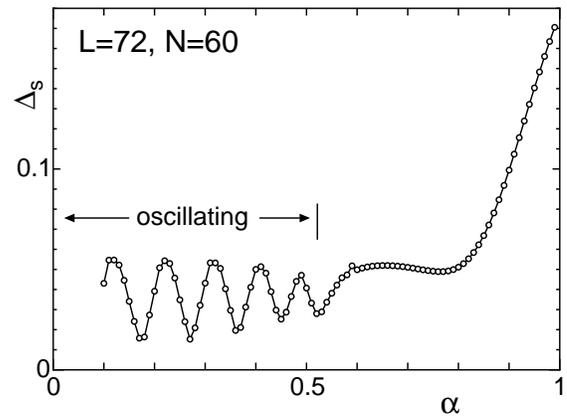,width=7.5cm}
\caption{$\alpha$-dependence of the spin gap $\Delta_s$ of the zigzag Hubbard model for $L=72$ and $N=60(\rho=0.833)$.  The on-site Coulomb interaction is fixed at $U=8$.
In $\alpha>0.7$(corresponding to the dotted line in FIG.\ref{phase}), the system is in the spin gap phase.
In $\alpha < 0.5$, $\Delta_s$ shows the rapid oscillation with respect to $\alpha$.
}
\label{oscill}
\end{figure}

In the region ``oscillating'', the behaviors of the spin gap becomes subtle,  as in FIG.\ref{sgap}(b).
The oscillating behavior of $\Delta_s$ makes  precise analysis of the spin gap in the bulk limit difficult.
In the vicinity of $\rho_c$, however,  we have seen that the size dependence of $\Delta_s$ is not so large, suggesting the existence of the spin gap, which is relevant to Pr247(see Sec. V).
As $\rho$ increases, the irregular oscillation to $\rho$ becomes significant and the size extrapolation of $\Delta_s$ breaks down.
The upper bound of the oscillating region is indicated by the open triangles, which are corresponding to the weak kink in the $\mu$-$\rho$ curve.
In order to illustrate another aspect of the oscillating region, we further investigate $\Delta_s$ as a function of $\alpha$ for a fixed $\rho$.
Figure \ref{oscill} shows the $\alpha$-dependence of  $\Delta_s$ for $L=72$ and $N=60(\rho=0.833)$.
We can then find that $\Delta_s$ exhibits the rapid oscillation with respect to $\alpha$ for $\alpha < 0.5$, which suggests that the incommensurate nature due to the frustration and the boundary effect cooperatively induce such subtle behavior of $\Delta_s$.
The rapid increase of $\Delta_s$ for  $\alpha > 0.8$ corresponds to the spin gap phase in Fig. \ref{phase}.
In the intermediate region($0.5< \alpha < 0.7$) the size dependence of $\Delta_s$ suggests gapless or almost gapless spin excitation, as mentioned before.

Here we would like to make a comment on the resolution of the phase diagram within $L=72$;
For example we can not distinguish  whether the boundary between the gapless region and the oscillating region/spin gap phase(open triangles/dotted line in Fig.\ref{phase}) is a bulk phase transition or not, chiefly because of the limited system size.
Also such a narrow phase as C2S2 mentioned by the weak coupling theory\cite{fabrizio}  is beyond the resolution of the discretized density of the electron.
In addition, we note that the spin gap for $\alpha <0.3$ is not evaluated properly  within the DMRG calculation, since  the  energy scale of the spin gap itself  becomes too small in the decoupled chains limit.

\section{discussions}

On the basis of the phase diagram, let us discuss the filling dependence of Pr247.
As was discussed in Ref.\cite{sano}, the chemical potential shifts from the quarter filling toward the electron doping side by oxygen reduction.
{\it If assuming} that $T_c$ is proportional to the amplitude of the  spin gap in the zigzag ladder, we can see that the effective model parameter of Pr247 is most likely in $\alpha\simeq 0.5 \sim 0.7$, for which the $\delta$ dependence of $T_c$  basically agrees with the filling dependence of the spin gap\cite{yamada}.
Thus, Pr247 may be located near the most competing region in the  phase diagram where the spin gap shows the subtle behavior, which is consistent with the relatively low $T_c$ of Pr247.
In addition, it is suggested that  Pr247 is sensitive to modification of  parameters due to external fields such as high pressure effect.\cite{fukuda}
As confirmed in the NQR/NMR experiments\cite{shinji},  a good one-dimensionality can be expected for Pr247 due to the superstructure of the metallic zigzag ladders,  the insulating single chains and  CuO$_2$ planes.
Such superstructure might be related to the relatively large $\alpha$, compared with a similar ladder compound YBa$_2$Cu$_4$O$_8$.
Very recently, Nakano {\it et al} have made FLEX calculation on the basis of the band structure of YBa$_2$Cu$_4$O$_8$, which nevertheless suggests the $s$-wave superconductivity in the small $\alpha$ region\cite{nakano}.
Thus the consistency with the actual band structure of Pr247, which is not available now experimentally and theoretically, is highly interesting.

In this paper, we have clarified  the various interesting behaviors induced by the frustration effect in the phase diagram.
In particular, the enhancement of the spin gap above $\rho_c$ and  the appearance of the oscillating region may be essential in  the connection with the Pr247 experiments.
However, we did not mention the correlation functions and the pairing symmetry here, which has been still unknown experimentally. 
In order to determine the paring symmetry,  it is needed to precisely investigate the correlation functions for a sufficient long chain.
Also the $U$-dependence of the zigzag Hubbard model, e.g. the  connections to the t-J model and the weak coupling theory, is a theoretically important problem.
We hope the present work to be a portal of further researches on the Pr247 superconductivity from both theoretical and experimental view points.

\acknowledgments

I thank to Y. Yamada, Y. \=Ono, K. Sano and T. Hikihara for fruitful discussions.
This work is partially supported by Grants-in-Aids for Scientific Research(No.18740230,17340100) and Grant-in-Aids for Scientific Research on Priority area(No.451,436) from MEXT.
Numerical computations were mainly performed on SX8 at Yukawa Institute in Kyoto University.


\end{document}